**Sharp resonant multiplet in femto-second optical pair-breaking spectroscopy of $YBa_2Cu_3O_{7-\delta}$**

Eric Li[*], J. J. Li[†], R. P. Sharma[*], S. B. Ogale[*], W. L. Cao[†], Y. G. Zhao[*,‡], C. H. Lee[†] and T. Venkatesan[*,†]

[*]*Center for Superconductivity Research and NSF MRSEC on Oxide Thin Films and Surfaces, Department of Physics, University of Maryland at College Park, College Park, MD 20742.*

[†]*Department of Electrical Engineering, University of Maryland at College Park, College Park, MD 20742.*

**Abstract**

**Femto-second optical pair breaking spectroscopy is performed on optimally doped $YBa_2Cu_3O_{7-\delta}$ near 1.5eV. Dramatic photon energy dependence with a sharp triplet fine structure is seen for the first time. The peak separations are attributable to the reported phonon and magnetic excitations in the system. The power dependence is non-linear with differing exponents for the three contributions. These results imply the presence of insulating antiferromagnetic domains in the superconducting state of YBCO over sub-picosecond time scale, and thereby favor an electronic phase separation picture.**

Since the discovery of high $T_C$ superconductivity (HTS) in 1986, several attempts have been and are still being made to seek a theory for this phenomenon in terms of the Fermi Liquid picture, by suitably flexing the conventional picture to accommodate the characteristics that are unique to the cuprate systems. Experimental results have, however, continued to defy and challenge the norms laid out by such an established paradigm, and there is growing evidence to suggest that the presumption of a featureless quantum gas or liquid of quasi-particles may in fact be too naive and simplistic for these systems[1]. One physical picture which attempts to capture the essence of the existence of a complex quantum matter in high $T_C$ cuprates, as indicated by several recent experiments, is the electronic phase separation (EPS) model; the so-called "stripe phase" scenario being one of its structured manifestations represented by a self-assembled array of conducting stripes separating hole-free antiferromagnetic (AFM) insulating domains. The foundations of these concepts were laid by the works of Zaanen *et al.*[2], Emery *et al.*[3,4], Schulz[5] and White *et al.*[6]. Experimentally there is growing evidence for the existence of complex textures of charge and spin in HTS, especially in the $La_2CuO_4$ family of superconductors, wherein the stripe dynamics are expected to be slow or suppressed[7-10]. However, the corresponding evidence is less direct in other key HTS materials such as $YBa_2Cu_3O_{7-\delta}$ (YBCO) and $Bi_2Sr_2CaCu_2O_{8+\delta}$ (BSCCO) where the stripe dynamics are suggested to be fast, although some recent experiments do suggest stripe-like but more disordered and dynamic fluctuations in these systems[11-14].

A key approach to understanding superconductivity is to probe the superconducting gap function $\Delta(\omega, \mathbf{k}, T)$ by inducing excitations in the system, and examining the corresponding quasi-particle behavior. Tunneling and inelastic neutron scattering studies focus on low energy excitations, while optical experiments generally employ high excitation energies. Holcomb *et al.*[15] used thermal-difference reflectance (TDR) spectroscopy to obtain the superconducting to normal reflectance ratio, $R_S/R_N$, for energies up to ~ 5eV in different superconducting cuprates, and found considerable deviations of this ratio from unity for photon energies near 1.5eV. They argued that these optical structures are hard to understand unless the electron-boson coupling function is assumed to consist of both a low-energy component (< 0.1eV) and a high-energy component located around 1.5eV. Stevens *et al.*[16] corroborated this claim by using femto-second time-resolved pump-probe spectroscopy. In both these experiments however, the measured physical quantities were the changes in the dielectric response, the correlation to electron pairing being only indirect. In this work, we selectively and directly measure the Cooper pair breaking rate via an *electrical* measurement[17] on optimally doped YBCO under femto-second laser excitation. For this temporal condition, the dynamic stripes, if present, would appear frozen, thereby exposing both the insulating and superconducting regions to the optical absorption process. The ability to selectively filter out the pair breaking contribution from a host of other possible excitation effects afforded by this scheme (signal at a few parts in 10) distinguishes it clearly from other optical experiments, including the pump-probe ones, where the measurement channel is also optical (signal at a few parts in $10^4$). Our data reveal for the first time, a sharp resonant multiplet in YBCO, which, in our view, strongly favors the electronic phase separation scenario.

The YBCO thin films were prepared by pulsed laser deposition on (100) oriented $LaAlO_3$ single crystal substrates. The film thickness was 100nm, nearly equal to the penetration depth of the laser beam at wavelength centered around 810nm[18]. The $T_c$ was ~90K and $J_c$ was $>10^6 A/cm^2$ at 77K. The films were patterned to obtain coplanar waveguide structures. The experimental set up and the device schematic are shown in Fig. 1. The size of the bridge at the center of the device was 5mm X 30μm. The device was mounted on a cold finger located in a cryogenic system (vacuum $<10^{-6}$ Torr), where the substrate temperature could be controlled between 10 and 300K with ±0.1K stability. The dc bias current was 4mA. The device was illuminated with laser pulses from a Ti:sapphire laser system, consisting of an Argon ion pump laser, an oscillator and a regenerative amplifier, generating 100fs laser pulses at up to 5µJ/pulse. The repetition rate was chosen to be 10kHz, which eliminates accumulation of prior pulse effects, and leads only to a fast optical response (FOR) with rise and fall times of the order of picosecond[17]. The typical laser fluence is 10µJ/cm$^2$/pulse. The wavelength of the laser was tunable within the range of 750nm – 850nm (1.65eV – 1.45eV). When the laser pulses illuminate the bridge, a transient drop occurs in the current flowing across the device. The corresponding waveforms were monitored by a fast digital sampling oscilloscope with a temporal resolution of 20ps.

A typical waveform of the FOR signal is also shown in Fig.1. The strongest peak is the primary signal, while the weaker peaks are due to reflections off the impedance mismatch on the transmission line. Two mechanisms have been put forward to explain the FOR signal, the kinetic inductance model[19,20] and the photo-activated flux flow model[21]. However, the kinetic inductance model has gained favor since it was shown that the amplitude of the FOR signal was not affected by applied magnetic field[19] in contradiction to the prediction of the photo-activated flux flow model. Moreover, the amplitude of the FOR signal shows a linear dc bias current dependence which is expected for the kinetic inductance model[19,20], whereas the photo-activated flux flow model predicts a quadratic dependence. Using the kinetic inductance model, the amplitude of the FOR signal can be shown to be approximately proportional to the time derivative of the kinetic inductance ($L_{kin}$) of the device, FOR $\propto \Delta L_{kin}/\Delta t$. $\Delta L_{kin}/\Delta t$ is related to the Cooper pair breaking rate by $\Delta L_{kin}/\Delta t = (m^*l/e^{*2}wdn_{sc}^2)(\Delta n_{sc}/\Delta t)$, where $m^*$ and $e^*$ are the effective mass and the effective charge of the Cooper pairs, $n_{sc}$ is the Cooper pair density, $\Delta n_{sc}/\Delta t$ is the Cooper pair breaking rate, l, w and d are length, width and thickness of the superconducting bridge, respectively. Thus, the Cooper pair breaking rate is measured directly and selectively in our experiment; all the other processes which do not break pairs being excluded. This indeed is the unique and distinguishing feature of the experiment, since it can probe the estimated less than 1% fraction of the electromagnetic energy in the laser pulse contributing to pair breaking[17], which could be easily missed by other approaches.

The photon energy dependence of the FOR signal at three different temperatures is shown in Fig. 2. The error bars were determined from five repeated measurements at each photon energy, and for each measurement, a minimum of 1024 waveforms were averaged for good statistics. Three sharp resonance peaks as indicated by A, B and C can be easily identified in the spectra at each substrate temperature. The solid lines were obtained by fitting the data to a theoretical model of summation of three Lorentzian line-shape functions, the dotted lines being the individual Lorentzian functions. At T = 80K, the peak positions are $E_A$ = 1.62eV, $E_B$ = 1.54eV and $E_C$ = 1.50eV. At T = 60K, the positions of all the peaks remain almost the same within experimental resolution of ~ 10meV due to the finite, Fourier-transform-limited width of the laser pulse, but the relative spectral weights exhibit changes. In fact, the contribution of C (A) grows (diminishes) with respect to that of B. At 20 K, the positions of peaks B and C still hold, with C increasing even further relative to B; while no clear signature of peak A is seen. It is interesting to note that the difference between peak B and peak C remains ~ 40meV for all three temperatures, which is curiously close to the famous 41meV peak observed in neutron scattering experiment on YBCO systems[22,23]. Interestingly, the difference between peak A and peak B is ~75meV, which is close to the energy of longitudinal optical (LO) oxygen bond-stretching phonons (~70-80meV) in the $CuO_2$ plane observed in inelastic neutron scattering measurements on HTS systems[24]. The reported zero momentum energy separation between the optical and acoustic magnon branches is also close to ~ 80meV in $YBa_2Cu_3O_{6.0+x}$ systems[25]. However, since the LO phonon mode and the 41meV magnetic excitation are the only sharp collective modes in the system, they appear to be more relevant to our case.

The most striking feature of these results is the extremely sharp spectral width of the resonance, ~100meV overall and ~20-50meV for the fine structures. This is at least a factor of 5-10 smaller than the minimum width expected for an electronically homogeneous conventional metallic or superconducting state, where the electronic bandwidths are several hundred meV. On the other hand, such sharp linewidths are reported for excitonic transitions in semiconductors and insulators. These two factors together imply that the femtosecond pulse appears to encounter some insulating regions in the superconducting state in the YBCO films, and that the absorption in these regions influences the pair breaking process. This latter implication is attributed to the fact emphasized earlier, namely, that in our experiment we ultimately and selectively record only the pair breaking processes. This understanding is completely consistent with the so-called stripe phase scenario, which suggests a self-organization of doped holes in the form of conducting stripes, separated by insulating AFM regions. Although these stripes are suggested to be highly dynamic in YBCO and therefore invisible for slow probes, the femto-second excitation pulse should see a snapshot of the dynamics and therefore the insulating regions.

If we presume, for the sake of argument, that we are indeed looking at frozen picture of stripes, we recognize that the insulating regions are not simply important from their electronic property perspective, but also from the magnetic one. In fact they support an AFM order, which is considered to be responsible for expelling the holes in the first place. Thus, any perturbation to the attendant driving forces leading to self-organization, electrical or magnetic, can be expected to have consequences for pair breaking. In the light of this expectation it is heartening to see that our spectral features exhibit energy separations which have been identified in the literature with magnetic and phonon excitations in the system: ~40meV corresponding to the magnetic resonance peak observed in neutron scattering experiments on YBCO[22,23], and ~75meV, the LO phonon energy scale observed in both neutron scattering measurements[24] and high-resolution angle-resolved photoemission (ARPES) measurements[26]. It has been suggested[24,26] that the LO phonons couple strongly with doped charges and therefore contribute significantly to the pairing process. Hence it is not surprising to see this energy scale appearing in our pair-breaking experiment.

As emphasized earlier, in the conventional optical experiments on optimally doped YBCO[27], no sharp resonance feature is seen for photon energies ~ 1.5eV. However, with enhanced oxygen under-doping and elimination of the metallic character, a resonance feature does appear at ~ 1.7eV. This has been identified as the $d^9$–$d^{10}\underline{L}$ Cu-O charge transfer (CT) exciton along the copper-oxygen planes. The close energy proximity of this CT exciton to the position of peak A in our data (1.63eV) tempts us to associate peak A with this CT exciton. Since our snapshot picture unveils the insulating regions to the exciting photon field, the CT peak could occur in such a picture. The energy could however get renormalized to some extent due to the electrostatic polarization effects attributable to charge separation over nanometer scale in the phase separation scenario. Here, it is useful to point out that a similar experiment performed on epitaxial, optimally doped superconducting $La_{1.85}Sr_{0.15}CuO_4$ (LSCO) thin films did not show any resonance in the FOR[28]. This is indeed as per our expectation, since the CT exciton in LSCO is near 2eV, which is out of the energy range examined in our experiment. This not only shows that the behavior seen by us in YBCO is intrinsic to the system, but also reinforces our assignment of peak A to CT exciton in insulating YBCO. With this assignment, one can then identify the other two peaks B and C with compound excitations involving the CT exciton and one, and two other excitations of phonon and magnetic origin, indicated earlier. The CT exciton in itself causes local spin rearrangement in the Cu-O bond, and thereby a ferromagnetic fluctuation in an AFM background, which can break the pairs, or influence the superfluid phase stiffness[4].

We now recall that the contribution of peak A is smallest at 80 K and decreases further with decreasing temperature, while the contribution of peak C involving the important ~40meV magnetic excitation and the ~75meV LO phonon grows with decreasing temperature with reference to the peak B. Given the fact that our measurement selectively probes only the pair breaking processes, the spectral intensity distribution and its evolution with temperature suggests a rather strong connection of the pairing to the magnetic excitations and the phonons, and possibly a rather indirect connection to the purely optical CT exciton. The fact that the CT exciton leads to a certain degree of pair breaking as observed in our experiment (presuming the assignment of peak A with such exciton) can be traced to the weakening of the phase stiffness within the stripe phase model[4]. Indeed, the notion of the change of kinetic inductance could have an entirely new microscopic interpretation in this scenario.

It is now useful to point out that several electronic processes in solids which can be identified with the coupling of charges with the lattice and spin systems are known to be characteristically non-linear[29,30]. Interactivity of extended or non-local fluctuations in the system heightens the non-linearity. Thus, based on the above considerations pertaining to the CT (A) and CT-coupled-magnetic-phonon excitations (B, C) suggested by our data, one should expect a nonlinear dependence of these contributions on photon density. Moreover, since B and C involve coupling of two and three excitations, respectively, they should, in fact, be increasingly nonlinear. This is precisely what is observed, as shown in Fig. 3, which gives the laser power dependence of the FOR signals corresponding to peaks A, B and C, for the measurement at 80K. A fit of these data (solid lines) to a simple power law [FOR ~ $P^b$ where P is the average laser power] gives the exponent b = 1.58±0.04, 1.86±0.05 and 2.29±0.14, for peaks features A, B and C respectively.

In conclusion, a pair breaking spectroscopy study using 100fs laser pulses performed on optimally doped, epitaxial YBCO films in their superconducting state, reveals a dramatic photon energy dependence near 1.5eV, with a sharp triplet fine structure. The laser power dependence of the triplet shows a non-linear behavior, with higher non-linearity for the peaks at lower energies. The narrowness of line-widths, the appearance of CT exciton albeit with renormalized energy, and the specificity of peak separations attributable to the reported magnetic excitation and phonon energies, suggest the existence of insulating AFM domains in the superconducting state of YBCO over the picosecond time scale. This is consistent with the electronic phase separation picture of high temperature superconductivity.

The authors would like to thank C. Lobb for reading of the manuscript and suggestions. This work was supported by the Office of Naval Research under Grant No. ONR-N000149611026 (Program Monitor: Dr. Deborah Van Vechten) and University of Maryland NSF-MRSEC under Grant No. DMR 00-80008.

‡ Present address: Department of Physics, Tsinghua University, Beijing 100084, People's Republic of China.

Figure Legends

Figure 1:   Experimental setup, device schematic and typical waveform for a fast optical response signal.

Figure 2:   Fast optical response as a function of photon energy, at three different temperatures below $T_C$.

Figure 3:   Fast optical response as a function of the average laser power, for the three individual contributions to the multiplet.

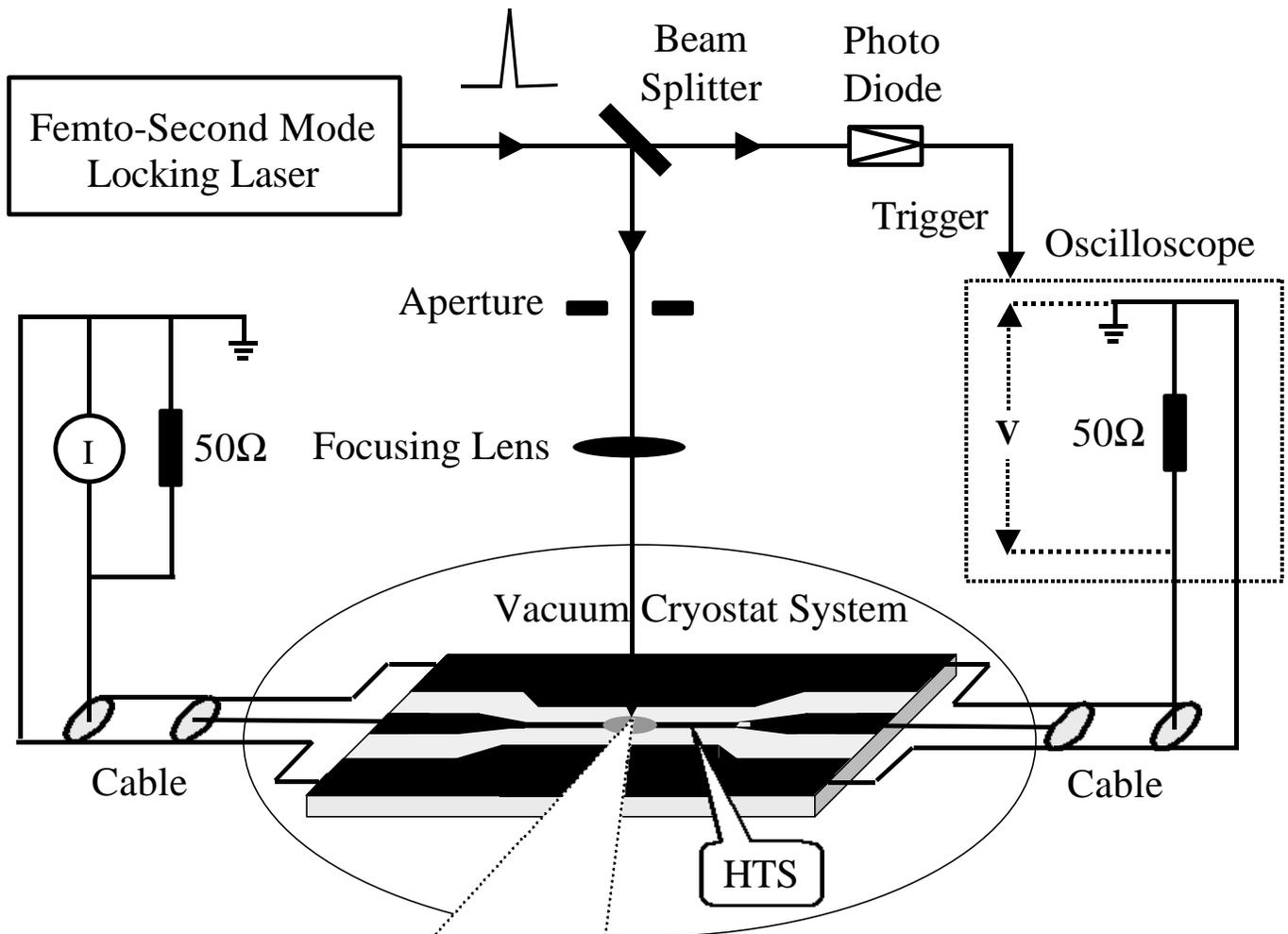
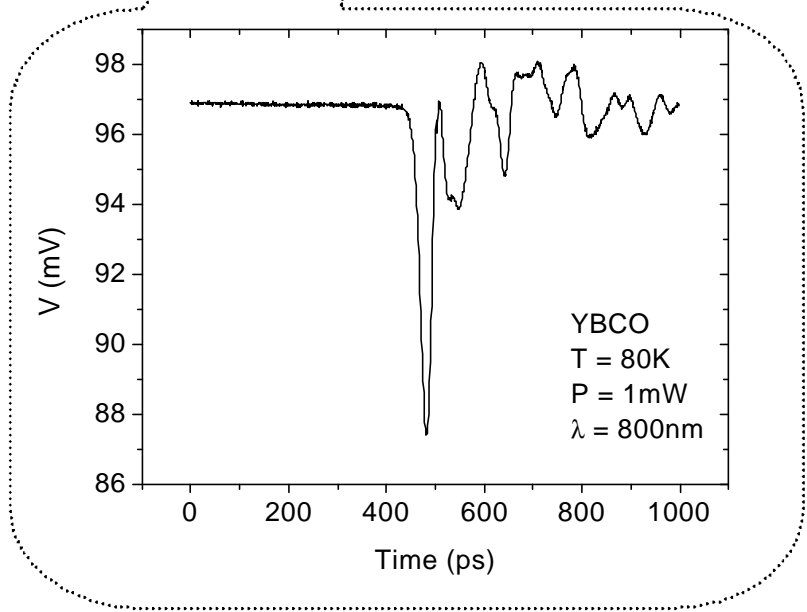

YBCO
T = 80K
P = 1mW
λ = 800nm

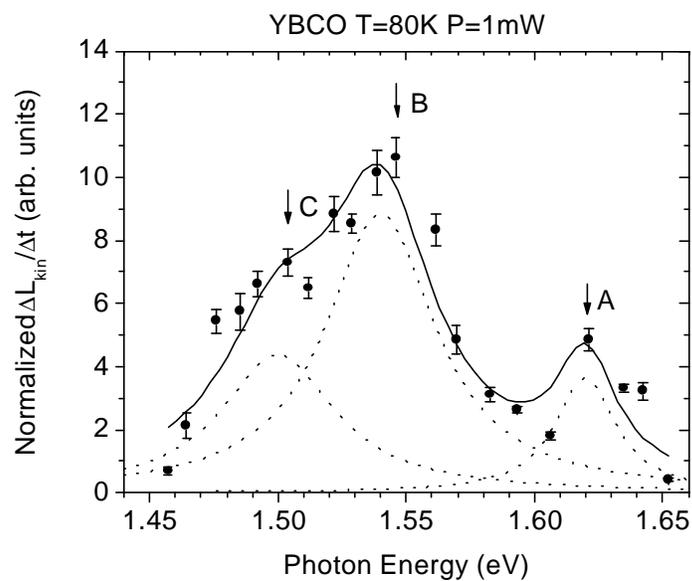
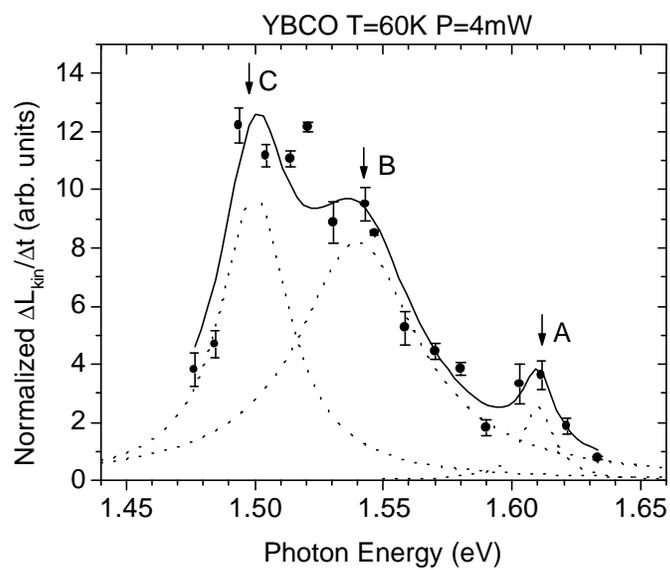
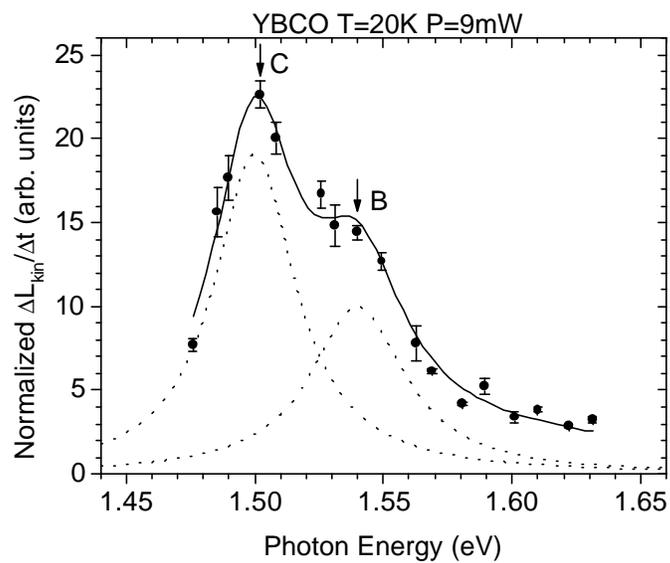

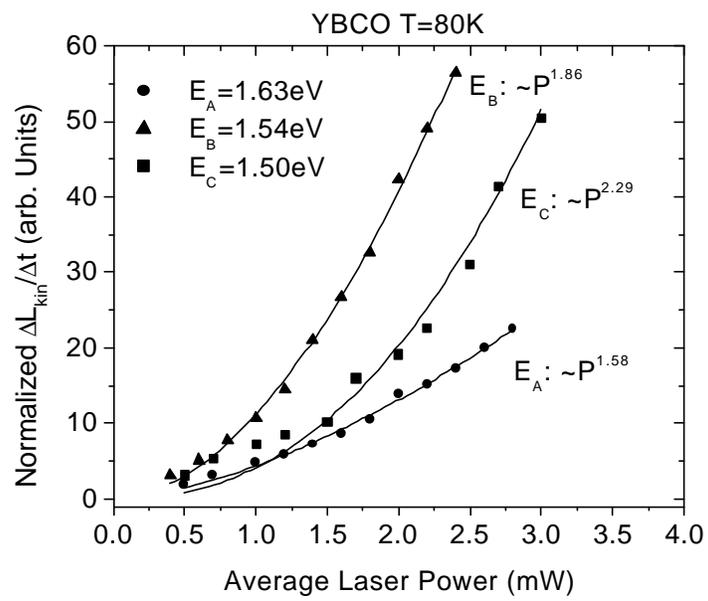